\documentclass[aps,twocolumn,groupedaddress,showpacs]{revtex4-1}
\usepackage{graphicx}% Include figure files
\begin{document}

\title{Center motions of nonoverlapping condensates coupled by long-range
dipolar interaction in bilayer and multilayer stacks}

\author{Chao-Chun Huang and Wen-Chin Wu}

\affiliation{Department of Physics, National Taiwan Normal
University, Taipei 11650, Taiwan}

\begin{abstract}
We investigate the effect of anisotropic and long-range dipole-dipole
interaction (DDI) on the center motions of
nonoverlapping Bose-Einstein condensates (BEC) in
bilayer and multilayer stacks. In the bilayer, it is
shown analytically that while DDI plays no role
in the in-phase modes of center motions of condensates,
out-of-phase mode frequency ($\omega_o$)
depends crucially on the strength of DDI ($a_d$). At the small-$a_d$ limit,
$\omega_o^2(a_d)-\omega_o^2(0)\propto a_d$.
In the multilayer stack, transverse modes associated with
center motions of coupled condensates are found to be optical phonon
like. At the long-wavelength limit, phonon velocity is proportional
to $\sqrt a_d$.
\end{abstract}

\maketitle

%%%%%%%%%%%%%%%%%%%%%%%%%%%%%%%%%%%
\section{Introduction}
In contrast to the isotropic $s$-wave contact interaction,
dipole-dipole interaction (DDI) has a very distinct
character -- anisotropic and long-range. This character
results in various interesting phenomena in ultracold dipolar
atom or molecule systems.
The studies of DDI effect on the structure and
dynamics of quantum many-body systems is at the forefront
of both theoretical and experimental interests.
In recent years dipolar chromium ($^{52}$Cr) atoms were created
with a magnetic moment of $6\mu_B$ ($\mu_B$ is
Bohr magneton), which is equivalent to a dipole moment
$d \approx 0.056 D$ ($1D\simeq 3.335 \times 10^{-30}
\textrm{C}\cdot \textrm{m}$)~\cite{griesmaier05}.
More recently, creations of dipolar molecules have also been achieved
in ultracold heteronuclear molecules such
as $^{40}$K$^{87}$Rb with a much stronger dipole moment $d\approx 0.6 D$
~\cite{ni08,ospelkaus08}. From the theoretical point of view,
many interesting physical properties have been studied, such as
biconcave shapes of the ground state~\cite{wilson08,ronen07,dutta07},
intriguing collapse mechanisms~\cite{parker09,koch08,lahaye08,ronen07,santos00},
vortex structures~\cite{wilson09,bijnen07,zhang05},
anisotropic solitons~\cite{tikhonenkov08} and so on.

Among many others, properties of
{\em nonoverlapping} dipolar Bose-Einstein condensates (BEC)
in bilayer or multilayer (quasi-1D optical lattice) stacks
have attracted great attention in recent years. The topics under investigation
include phonon instability~\cite{koberle09,klawunn09}, soliton-soliton
scattering~\cite{nath07}, pair superfluidity~\cite{arguelles07},
and filament condensation~\cite{wang06}.
In these systems, condensates are effectively nonoverlapping
between neighboring layers and as a matter of fact,
contact or even short-range interaction
plays no role between neighboring
layers. However, long-range DDI
will still play an important role across different layers.
Moreover in a real system, dipoles of the dipolar gas
are usually aligned (by applying strong electric or magnetic field)
along the direction of the stacks. This in turn will make dipolar molecular
systems more stable against recombination and result in better opportunity
of forming dipolar molecular BEC.

This paper attempts to study collective excitations of
nonoverlapping atomic or molecular BEC in bilayer and multilayer stacks, that are
coupled by long-range DDI. The focus will be placed on the motions of the
{\em centers} of condensates.
When weakly interacting Bose condensates are loaded into a single harmonic
(magnetic) trap, it is well known that, for small oscillations, center motion
of condensates will undergo
harmonic oscillation of frequency equal to the characteristic
trap frequency~\cite{garcia96,garcia97}.
This remains true even when the system has a long-range DDI in it.
When condensates are loaded into a double well and form a bilayer
system in, say, $z$ direction (while a single harmonic trap is applied in
the $x$-$y$ plane),
center motions of the two condensate clouds
can be very interesting upon the activation of long-range DDI.
It will be shown explicitly that
when DDI is present, both in-phase and out-of-phase motions of the
two condensate centers will still undergo harmonic oscillations
(for small oscillations). However, while in-phase modes
are independent of DDI, out-of-phase modes
will depend crucially on the strength of DDI.

In addition, it is also interesting to investigate
center motions of nonoverlapping condensates in a multilayer stack
(quasi-1D optical lattice along $z$-direction).
By properly treating the boundary effect (see later),
transverse modes associated with motions of the centers of condensates
in each layer are found to be optical phonon like.
At infinitely long wavelength limit ($q_z\rightarrow 0$),
phonon mode frequency just equals to the harmonic trap frequency in the
$x$-$y$ plane and
at the long-wavelength limit ($q_z z_0\ll 1$, $z_0$ is the lattice constant),
phonon mode velocity is found to be proportional to the
square root of the strength of DDI.
%Here we assume the number of site is large enough large such that
%boundary effect can be ignored. In this kind system there is a
%important question which is how many neighbor sites, $N_c$,
%must be included in calculation.
Due to the long-range character of DDI, for the multilayer system of
finite number of layers,
it is important to treat properly the boundary effect. With this regard,
we have introduced a truncation number ($N_c^g$) which corresponds to
number of neighboring layers
included for satisfactorily converged results.
It is found that
$N_c^g$ will depend on the ratio of lattice constant ($z_0$) and
condensate radius in the $x$-$y$ plane only.

To end this introductory section, we emphasize two things.
Firstly, the results presented in this paper, especially the dependence of
DDI strength on center motion mode frequencies, are believed to be valid
even when the system is not Bose condensed.
Since long-range DDI survives in the non-condensed system,
it is encouraging that the experiment can also be done on the
ultracold heteronuclear dipolar molecules
which exhibits a strong dipole moment but nevertheless is yet to be
Bose condensed ~\cite{ni08,ospelkaus08}. Secondly,
the strength of DDI can actually be extracted through
measurements of out-of-phase mode in the bilayer or transverse phonon mode
in the multilayer stack.

The paper is organized as the following. In Sec.~\ref{sec2},
energy functional of a nonoverlapping bilayer system is
given. Proper trial wave functions are introduced within the
variational framework and minimized ground-state energies are obtained.
In Sec.~\ref{sec3},
analytical results of the mode frequency for center motions of
condensates are given. It is shown that
while in-phase mode
frequency is independent of the strength of DDI, out-of-phase modes
depend crucially on the strength of DDI.
In Sec.~\ref{sec4}, we extend the study to the center motions of condensates in
a multilayer stack.
By properly treating the boundary effect, transverse phonon modes
are obtained. At the long-wavelength limit, we show that
phonon velocity is proportional to the square root of the strength of DDI.
Sec.~\ref{sec5} is a conclusion.

\section{NONOVERLAPPING BILAYER SYSTEM} \label{sec2}

We consider a nonoverlapping bilayer system
with same BEC atoms or molecules in each layer.
Energy functional of the system is given by
\begin{eqnarray}
E&=&E_1+E_2+E_{12},
\label{eq:E}
\end{eqnarray}
where ($i=1,2$)
\begin{eqnarray}
&&E_i =  {{N}\int {d{\bf{r}}\psi _i^*({\bf{r}})\left[ {
- \frac{{{\hbar ^2}}}{{2m}}{\nabla ^2} + {V_{\rm
ext,i}}({\bf{r}}) } \right.} }  \label{eq:pseudoenergy1}
\\
&&+ \left. \frac{N-1}{2}\left( {g}|{\psi _i}({\bf{r}}){|^2}+
\int {d{\bf{r'}}{V_{dd}}({\bf{r}} -
{\bf{r'}}){{\left| {{\psi _i}({\bf{r^\prime}})} \right|}^2}}
\right)\right] {\psi _i}({\bf{r}})
\nonumber
\end{eqnarray}
and
\begin{eqnarray}
E_{12}&=&{N^2} \int \int{d{\bf{r}}d{\bf{r'}}{V_{dd}}({\bf{r}}
-{\bf{r'}}){{\left| {{\psi _1}({\bf{r}})} \right|}^2}{{\left|
{{\psi
_2}({\bf{r^\prime}})} \right|}^2}}.
\label{eq:pseudoenergy}
\end{eqnarray}
Here $E_1$ and $E_2$ correspond to intralayer energies and
$E_{12}$ is the interlayer energy caused by DDI.
$N$ is the number of condensed atoms or molecules in
each layer and $\psi_i$ is the normalized wave function for layer $i$.
$g=4\pi \hbar^2a/m$ with $a$ the $s$-wave scattering length and
$V_{dd}({\bf{r}})=d^2(1-3\cos^2\theta)/|\bf{r}|^3$ is the
dipole-dipole interaction with
$\theta$ the angle between $\bf{r}$ and the dipole orientation.
For magnetic dipoles, the strength
$d^2 =\mu_0\mu_m^2/4\pi$ with $\mu_m$ the magnetic dipole
moment, while for electric dipoles, $d^2 = d_e^2
/4\pi\varepsilon_0$ with $d_e$ the electric dipole moment.
For a nonoverlapping bilayer considered in current context,
condensate wave functions are not
overlapped across the two layers, {\em i.e.},
$\int d\mathbf{r}|\psi_1(\mathbf{r})|^2 |\psi_2(\mathbf{r})|^2\approx 0$.
As a consequence, interlayer energies associated with $s$-wave interaction
as well as interlayer hopping are neglected.

\begin{figure}[ptb]
\vspace{0.0cm}
\includegraphics[width=0.4\textwidth]{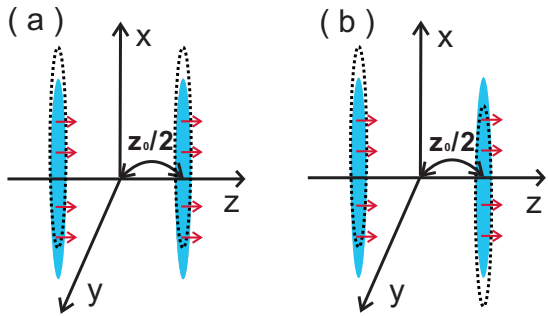}
\vspace{-0.2cm} \caption {Sketch of the nonoverlapping bilayer condensates
and their center motions along the $x$-direction.
Red arrows denote the direction of dipoles.
Blue stacks represent the stationary condensates,
while black dash lines represent their movements.
Frame (a) corresponds to the in-phase mode and
frame (b) corresponds to the out-of-phase mode.}
\label{fig1}
\end{figure}

To create a nonoverlapping bilayer ultracold BEC system,
one can set up a double well with a large barrier in the middle.
One possible candidate of nonoverlapping double potential well
in $z$ direction is $V=\frac{1}{2}m\left( {\omega _x^2{x^2} + \omega
_y^2{y^2}+ \omega _z^2{z^2}} \right) + {V_\ell }\cos ^2(
\pi z/d_\ell)$ by applying a harmonic
trap together with a deep optical (lattice) trap~\cite{albiez05}. Alternatively,
one can apply a harmonic
trap together with a repulsive barrier potential to obtain
$V= \frac{1}{2}m\left( {\omega _x^2{x^2} + \omega
_y^2{y^2} + \omega _z^2{z^2}} \right) + {V_0}\exp (
-z^2/\ell_z^2)$~\cite{andrews97}.
By controlling the values of $V_\ell$ {\em vs.} $d_\ell $ or
$V_0$ {\em vs.} $\ell_z$, effective nonoverlapping bilayer system can be
formed. As a matter of fact, wave function of each layer is extremely
narrow (pancake like) in $z$-direction and to the leading order,
trap potentials experienced by the atoms or molecules in each layer
can be approximated by the following anisotropic harmonic potentials,
\begin{eqnarray}
{V_{{\rm ext},{1,2} }}({\bf r}) = \frac{1}{2}m\left[ {\omega _{x}^2{x^2}
+ \omega
_{y}^2{y^2} + \omega _{z}^2{{(z \mp {z_1/2})}^2}} \right].
\label{eq:trap1}
\end{eqnarray}
Here $z_1$, determined by the experimental setup,
corresponds to the spacing between the two trap
minima located at $z=\pm z_1/2$. For nonoverlapping bilayer systems,
$\omega _{z}$ is typically much larger than $\omega _{x}$ and $\omega _{y}$.
In this paper, as depicted in Fig.~\ref{fig1},
we consider the dipolar system to which
dipoles are oriented along $z$ direction.
With this geometry,
the system will be the most stable compared to others.

Within the variational framework, Gaussian ansatz is used
for the trial wave functions of the bilayer system.
As long as the system is not close to a collapsed state,
Gaussian function should be a good trial wave function for the system
\cite{ronen06}. For layer $1$ and $2$, we then take
\begin{eqnarray}
{\psi_{1,2}}({\bf r}) = A \exp \left[ { - \frac{1}{2}\left(
{\frac{{{x^2}}}{{R_{x}^2}} + \frac{{{y^2}}}{{R_{y}^2}} +
\frac{{{{\left({z \mp {z_0/2}} \right)}^2}}}{{R_{z}^2}}} \right)}
\right],
\label{eq:trialfun}
\end{eqnarray}
where variational parameters $R_x$, $R_y$, and $R_z$ correspond
to condensate radii in $x$, $y$, and $z$ directions,
$A=1/ \sqrt{R_{x}R_{y}R_{z}\pi^{3/2}}$ is the normalization constant,
and $z_0$ corresponds to the distance between the two
condensate centers located at $z=\pm z_0/2$.
Note in general that $z_0$ can be different from $z_1$
[the latter corresponds to trap minima, see Eq.~(\ref{eq:trap1})].
When DDI vanishes, $z_0$ will be identical to $z_1$.
However, when DDI is present,
$z_0$ will be smaller (larger) than $z_1$ if
DDI is attractive (positive) in $z$ direction.
Nevertheless, for the present nonoverlapping bilayer system, $z_0$
should be not much different from $z_1$.

Truly speaking, due to anisotropic nature of DDI,
real wave function for each layer
will not be symmetric in $z$ direction even though a symmetric potential
trap~(\ref{eq:trap1}) is in effect.
However, when $\omega_z\gg \omega_x,\omega_y$,
wave function in each layer is still relatively symmetric
in $z$ direction.
Substituting trial wave functions~(\ref{eq:trialfun})
and trap potentials~(\ref{eq:trap1})
into energy functional~(\ref{eq:E})--(\ref{eq:pseudoenergy}), we obtain
\begin{eqnarray}
{E}&=&{N} \left\{ \frac{1}{2{R_{x}^2}} + \frac{1}{2{R_{y}^2}} +
\frac{1}{2R_{z}^2} +\lambda _{z}^2 \left[ \frac{R_{z}^2}{2} +
\frac{{\left( {{z_0} - {z_1}} \right)}^2}{4}\right]\right.
\nonumber\\
&+& \left. \frac{R_{x}^2}{2} +\lambda _{y}^2 \frac{R_{y}^2}{2}+{{N-1}\over R_{x}
R_{y} R_{z}}\left(\sqrt{2\over \pi}a_{s} - a_{d}F_0
\right) \right\} \nonumber\\
&-&{{N^2}\over R_{x}
R_{y} R_{z}}a_{d}F.
\label{eq:energyfun}
\end{eqnarray}
In (\ref{eq:energyfun}) and throughout this paper, we have rescaled the energy
$E/(\hbar\omega_{x} ) \rightarrow E$ and the time
$t\omega_x\rightarrow t$. Besides, all the lengthes
($R_x,R_y,R_z,z_0,z_1$) are
scaled by the magnetic length in $x$-direction, $\ell \equiv \sqrt
{{\hbar }/{m\omega_{x} }}$, and the ratios $\lambda_{y}\equiv
\omega_{y}/\omega_{x}$ and $\lambda_{z}\equiv
\omega_{z}/\omega_x$. Dimensionless coupling strength
$a_{d}\equiv d^2 m/\hbar^2 \ell$ and $a_s\equiv a/\ell$.
$F=F(z_0,R_x,R_y,R_z)$ is related to the interlayer energy due to DDI,
while $F_0\equiv F(z_0=0,R_x,R_y,R_z)$ is related to the intralayer energy due to DDI.
More explicitly, for dipoles aligned along the
$z$ direction, $F$ is given by the following integral
\begin{eqnarray}
 F &=& {1\over 6\pi^2} \int   {d{\bf{k}}
\exp\left[ { - \frac{1}{2}\left( {k_x^2 + k_y^2 + k_z^2}
\right)} \right]
 \cos\left({k_z} \frac{{z_0}}{R_z}\right) } \nonumber\\
&~&~~~~~~~~~~ \times\left(1- \frac{3k_z^2}{ k_x^2R_z^2/R_x^2
 +k_y^2R_z^2/R_y^2+k_z^2 }\right).
 \label{eq:twodipole2}
\end{eqnarray}
Values of $R_{x}$, $R_{y}$, $R_{z}$, and $z_0$ for the
ground state are obtained
by solving the conditions $\partial E/\partial R_{i}=0$ $(i=x,y,z)$ and
$\partial E/\partial z_0=0$.
%In the following section we use the energy functional
%to discuss the motion of the center in x direction.

\section{Center MOTIONS OF BILAYER SYSTEM}
\label{sec3}
As mentioned before, in this paper we focus on the
motions of the center of condensates along $x$ direction
(see Fig.~\ref{fig1}).
Generalization to considering as well the center motions of condensates along $z$
direction is straightforward.
However, for condensates strongly confined in $z$ direction, condensate
wave functions are rigid in $z$ direction
and as a consequence collective excitation in $z$ direction will cost more energy.
Besides observation of these motions of much smaller
amplitude are relatively more
difficult.

By variational approach,
suitable dynamical variables should be added into
the trial wave functions (\ref{eq:trialfun}). In
Ref.~\cite{garcia96,garcia97}, it was shown in
the single-trap system that equations of motion for
the center of condensates are decoupled from equations
of motion for the width of condensates.
For a bilayer system studied in this context, similar
decoupling occurs although the proof is omitted for brevity.
As a matter of fact, for the motions of the center of
condensates, dynamical wave functions can be taken to be
\begin{eqnarray}
{\psi_{1,2}}({\bf r},t) &=& {A}\exp \left[ - \frac{y^2}{2R_{y}^2} -
\frac{ (z\mp z_0/2)^2}{2R_{z}^2}\right]\nonumber\\
&~&\times \exp \left\{ { - \frac{\left[x - {x_{1,2}}(t)\right]^2}{2R_{x}^2} -
ix{c_1}(t)}  \right\},
 \label{eq:tfun}
\end{eqnarray}
where dynamical variables are added associated with the center motions only.
Here $x_i(t)$ corresponds to
the fluctuation of the center of condensate
of layer $i$ in $x$-direction, while
$c_i(t)$ corresponds to the sloping phases of the condensates of layer $i$.
The task is to find the equations of motion and solve it.

We start from the effective Lagrangian, $L=T+E$, where $T$ is given by
\begin{eqnarray}
T = \int {d{\bf{r}}\sum\limits_{j = 1,2} {{N}\left(
{\frac{{i\hbar }}{2}} \right)\left[ {{\psi _j}
\frac{{\partial \psi _j^*}}{{\partial t}} - \psi _j^*
\frac{{\partial {\psi _j}}}{{\partial t}}} \right]} }
\label{eq:langt}
\end{eqnarray}
and $E$ is given by Eqs.~(\ref{eq:E})--(\ref{eq:pseudoenergy}). Substituting
Eq.~(\ref{eq:tfun}) into the Lagrangian and expanding dynamical
variables up to second order (for small oscillations), one obtains
\begin{eqnarray}
L&=& \sum\limits_{i = 1,2} {N\left( { \frac{x_i^2}{2}
+\frac{c_i^2}{2}- {x_i}{{\dot c}_i}} \right)}
+{N^2 a_d G
(x_1-x_2)^2\over 2R_x^3 R_y R_z},
\label{eq:langsec}
\end{eqnarray}
where $G=G(z_0,R_x,R_y,R_z)$ and given by the integral
\begin{eqnarray}
 G&=& {1\over 6\pi^2}\int {d{\bf{k}}k_x^2
 \exp \left[ { - \frac{1}{2} \left( {k_x^2+k_y^2 + k_z^2}
\right)} \right]\cos \left({k_z}\frac{z_0}{R_z}\right) } \nonumber\\
&~&~~~~~~~~~~ \times\left( 1-\frac{3k_z^2}{ k_x^2R_z^2/R_x^2
 +k_y^2R_z^2/R_y^2+k_z^2 }\right).
\label{eq:dipoleper}
\end{eqnarray}
It is interesting to note that $s$-wave scattering coupling
$a_s$ has completely dropped out in $L$ in (\ref{eq:langsec}).
Equations of motion can be derived using the Lagrange equation,
$\frac{d}{dt}\frac{\partial L}{\partial \dot{q}}=\frac{\partial
L}{\partial q}$, where $q$ can be any one of the four dynamical variables.
By assuming $q=q_0\exp(i\omega t)$, after some algebra
we obtain two branches of excitation modes:
\begin{eqnarray}
{\omega _{i}^2} &=& 1,\nonumber\\
{\omega _{o}^2} &=&
1 + N a_d \left({2G\over R_x^3 R_y R_z}\right),
\label{eq:discenter}
\end{eqnarray}
where $\omega _{i}$ corresponds to the in-phase mode associated
with the solutions of $x_1=x_2$ and $c_1=c_2$
[such as the motion depicted in Fig.~\ref{fig1}(a)] .
While $\omega _{o}$ corresponds to the out-of-phase mode
associated with the solutions of $x_1=-x_2$ and $c_1=-c_2$
[such as the motion depicted in Fig.~\ref{fig1}(b)].

Since $x$-direction harmonic trap frequency $\omega_x$ is the energy unit
used in this context,
the results in (\ref{eq:discenter}) show that in-phase mode frequency of
$x$-direction center motions of bilayer condensates is just $\omega_x$,
independent of the long-range DDI.
For the out-of-phase mode, in contrast, the mode frequency is
shifted from $\omega_x$ and the deviation depends crucially on
the strength of DDI ($a_d$). To be more explicitly, the deviation
will depend on $a_d$, the number $N$, as well as the four lengths
($R_x$,$R_y$,$R_z$,$z_0$).
Since number $N$ and the four lengths are measurable quantities,
one can then calculate $G$ [using Eq.~(\ref{eq:dipoleper})] with
the measured values of $N$ and ($R_x$,$R_y$,$R_z$,$z_0$).
This simply means that once center motions of the bilayer is performed
and out-of-phase mode frequency
$\omega_o$ is measured, one can actually extract the value of
$a_d$ via Eq.~(\ref{eq:discenter}).

\begin{figure}[ptb]
\vspace{-0.3cm}
\includegraphics[width=0.45\textwidth]{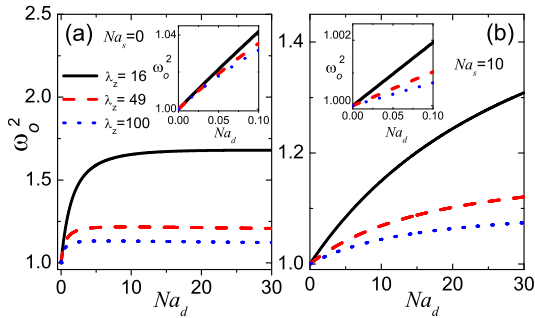}
\vspace{-0.4cm} \caption {The square of out-phase mode, $\omega_{o}^2$,
plotted as the function of $Na_d$ ($N$ is particle number in each layer
and $a_d$ is DDI coupling) for $\lambda_z
=16,~49,$ and $100$ respectively. The insets
show linear dependence of $Na_d$ at small $Na_d$.
In frame~(a), $s$-wave coupling $Na_s=0$ and
in frame~(b), $Na_s=10$. $z_1$ is  chosen to be $1.2\ell$ in both frames.}
\label{fig2}
\end{figure}

Fig.~\ref{fig2} shows the square of out-of-phase mode $\omega_{o}^2$
as the function of $Na_d$. For convenience, we consider
the isotropic case in $xy$ plane ($R_x=R_y\equiv R$ and
$\lambda_y=1$), and have chosen $z_1=1.2\ell$.
Fig.~\ref{fig2}(a) plots the $Na_s=0$ case with $\lambda_z$ chosen to be
$16$, $49$, and $100$ respectively. While Fig.~\ref{fig2}(b) plots
$Na_s=10$ case with the same choice of the three $\lambda_z$'s.
The insets
show the linear dependence of $Na_d$ when $Na_d$ is small.
Within variational framework, values of $R$, $R_z$,
and $z_0$ are determined by minimizing the energy functional~(\ref{eq:energyfun}).
It is useful to check that when both $s$-wave and DDI couplings are absent, $a_s=a_d=0$,
one finds that $z_0=z_1$, $R=1\ell$, and $R_z=(1/4)\ell, (1/7)\ell,$ and $(1/10)\ell$
corresponding respectively to $\lambda_z=16, 49$ and $100$ cases. However,
when $a_d$ is present,
$z_0$ becomes shorter than $z_1$, while $R$ and $R_z$
become larger than those for the case of $a_d=0$.
This simply means that the ratio of $z_0/R_z$
will become smaller in the presence of $a_d$.
Nevertheless, in the strong confining regime ($z_0/R_z\geq 4$)
considered in this context, the change of $z_0/R_z$ due to $a_d$
is minor. For example, in case of $a_s=0$ and
$Na_d=30$, ratio $z_0/R_z$ is found to be 4.07, 7.9, and 11.6
for $\lambda_z=16, 49$ and $100$, as compared to 4.8, 8.4, and 12 for
the case of $a_s=a_d=0$.

When results of Fig.~\ref{fig2}(b) are compared to those in
Fig.~\ref{fig2}(a), one sees that repulsive
$s$-wave coupling $a_s$ acts to reduce the deviation of
$\omega_{o}$ from $\omega_x$. While short-range $a_s$ plays no role between neighboring
layers, its repulsion
actually increases the radii of condensate ($R_x$, $R_y$, $R_z$) in each layer
and consequently
$G/R_x^3 R_y R_z$ becomes smaller.
This indicates that $\omega_o^2$ or its slope against $Na_d$ (at small $Na_d$)
will be smaller.  In addition, it is found that regardless of the value of $a_s$,
for the same $Na_d$, the larger $\lambda_z$ is, the closer $\omega_o$ is
to $\omega_x$. In $^{52}$Cr atom
dipolar BEC, it has been measured that $d^2m/ \hbar^2\simeq 24$
\AA. If atom number in one layer is $N \sim 10^4$
and the harmonic oscillator length $\ell \sim 1~\mu m$, then it is estimated that
$Na_d\sim 20$. This gives a reference how large
$\omega_o$ is when the results of Fig.~\ref{fig2} are considered.

\section{Nonoverlapping MULTILYAER STACK}\label{sec4}

In this section, we extend to study center
motions of condensates in a multilayer stack.
As before, condensates are assumed to be non-overlapping between
neighboring layers.
The kind of system can be realized in a deep quasi-1D optical
lattice~\cite{hadzibabic06}.
In an analogous way, energy functional of the multilayer system can be given by
\begin{eqnarray}
E&=&\sum_{n=1}^{N_s} \left( E_{n}+\sum_{m}E_{n,m} \right),
\label{eq:energyfunopt0}
\end{eqnarray}
where
\begin{eqnarray}
&&E_n =  {{N}\int {d{\bf{r}}\psi _n^*({\bf{r}})\left[ {
- \frac{{{\hbar ^2}}}{{2m}}{\nabla ^2} + {V_{\rm
ext}}({\bf{r}}) } \right.} } \label{eq:energyfunopt1}
\\
&&+ \left. \frac{N-1}{2}\left( {g}|{\psi _n}({\bf{r}}){|^2}+
\int {d{\bf{r'}}{V_{dd}}({\bf{r}} -
{\bf{r'}}){{\left| {{\psi _n}({\bf{r^\prime}})} \right|}^2}}
\right)\right] {\psi _n}({\bf{r}})
\nonumber
\end{eqnarray}
and
\begin{eqnarray}
E_{n,m}= N^2\int\int {d{\bf{r}}d{\bf{r'}}{V_{dd}}({\bf{r}}
-{\bf{r'}}){{\left| {{\psi _n}({\bf{r}})} \right|}^2}{{\left|
{{\psi_{n+m}}({\bf{r^\prime}})} \right|}^2}}.~~~
\label{eq:energyfunopt}
\end{eqnarray}
Here $N_s$ corresponds to the number of layers in the stack
and $N$ corresponds to the
number of atoms in each layer. $E_n$ represents
the intralayer energy for layer $n$, while $E_{n,m}$ represents the interlayer
energy between layers $n$ and $(n+m)$ coupled by DDI.
The external trap, $V_{\rm ext}$, consists of magnetic and
optical traps, where
magnetic trap is $m(\omega_x^2 x^2+\omega_y^2y^2)/2$,
while optical trap is $sE_r\sin^2(\pi z/z_0)$ with $z_0$ the spacing between
neighboring layers, $s$ the strength of optical trap, and
$E_r=\hbar^2\pi^2/2mz_0^2$ the recoil energy.

In the calculation, we shall use the following approximation for
$E$ in Eq.~(\ref{eq:energyfunopt0}):
\begin{eqnarray}
E
%\sum_{n=1}^{N_s} \left( E_{n}+\sum_{m}E_{n,m} \right)
\simeq\sum_{n=1}^{N_s} \left( E_{n}+\sum_{|m|=1}^{N_c}E_{n,m}
\right),
\label{eq:energyfunoptapp}
\end{eqnarray}
where $N_c$ is a truncation number to which how many neighboring sites are
included for $E_{n,m}$ ($|m|=1$ correspond to the two nearest neighbors).
With (\ref{eq:energyfunoptapp}),
periodic boundary condition (PBC) which connects the two ends is applied.
Owing to the long-range nature of DDI, it is expected that
for a satisfactory converged result, $N_c\gg 1$. On the other hand, for a trusty result,
it is required that the truncation number should be much smaller
than the total number of layers in the stack ($N_c\ll N_s$).
%It will be shown later that $N_c$ turns out to be
%the function of the ratio $z_0/R$ only.

\begin{figure}[ptb]
\vspace{0.0cm}
\includegraphics[width=0.4\textwidth]{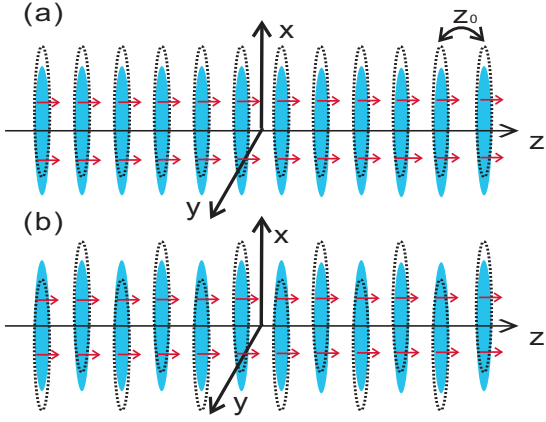}
\vspace{-0.1cm}
\caption {Sketch of the multilayer stack
and the motions of the center of condensates.
Frame (a) corresponds to a $q_z=0$ in-phase phonon mode and
frame (b) corresponds to a $q_z=\pi/z_0$ out-of-phase phonon mode.
Red arrows denote the direction of dipoles.
Blue stacks represent the stationary condensates and
black dash lines represent the motions of condensates.}
\label{fig3}
\end{figure}

In the current multilayer stack, for simplicity, we also apply
the Gaussian ansatz for the trial
wave function of each layer [similar to that of
Eq.~(\ref{eq:trialfun})].
It should be emphasized again that due to the boundary effect,
true wave function associated with each layer
is not perfectly symmetric in $z$ direction.
Nevertheless, for the current nonoverlapping condensates under study,
a symmetric wave function will be a good approximation for each layer.
After a lengthy derivation, we obtain the following
energy functional for the multilayer system
 \begin{eqnarray}
\frac{E}{N_s} &=& N\left\{ {\frac{1}{{4R_x^2}} +
\frac{1}{{4R_y^2}} + \frac{1}{{4R_z^2}} +
\frac{{s{E_r}}}{2}(1 - {e^{ - {\pi ^2}R_z^2/z_0^2}})}
\right.\nonumber \\
&+& \left.  \frac{{R_x^2}}{4}+ \lambda _y^2\frac{{R_y^2}}{4}
 + \frac{N-1}{2{R_x}{R_y}{R_z}}
 \left(\sqrt{2\over \pi} a_s- a_{d}F_{0}\right)  \right\}\nonumber\\
&-& \frac{N^2a_{d}}{2{R_x}{R_y}{R_z}} \Gamma(N_c),
\label{eq:energyfunopt1}
\end{eqnarray}
where
 \begin{eqnarray}
\Gamma(N_c)\equiv\sum_{|m|=1}^{N_c} {F_{m}}
\end{eqnarray}
with $F_{m}$ being just $F$ in Eq.~(\ref{eq:twodipole2})
with $\cos(k_zz_0/R_z)$ term replaced by $\cos(mk_z z_0/R_z)$. Once again,
it is assumed that dipoles of the dipolar gas are aligned along
the $z$-direction and we consider only the
motions of the center of condensates in $x$ direction for small
oscillations (see Fig.~\ref{fig3}).
In (\ref{eq:energyfunopt1}), all energies and lengths are rescaled
in the same way as those in the bilayer case.
%Besides, the ratios $\lambda_{y}\equiv
%\omega_{y}/\omega_{x}$ and $\lambda_{z}\equiv
%\omega_{z}/\omega_x$.
The recoil energy will then
reduce to $E_r=\pi^2\ell^2/2z_0^2$ in the dimensionless unit.

\begin{figure}[ptb]
\vspace{-0.0cm}
\includegraphics[width=0.4\textwidth]{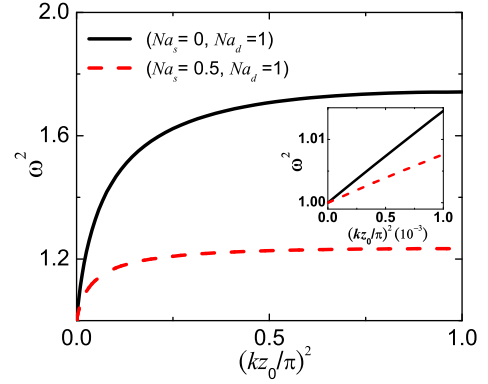}
\vspace{-0.4cm} \caption {Square of the transverse phonon modes plotted as
the function of $(kz_0/\pi)^2$. Two cases,
$(N a_s,N a_d)=(0,1)$ and $(0.5,1)$ are presented. The inset
shows the linearity at long wavelength limit ($kz_0\ll\pi$) and
the slope is equal to the square of phonon velocity.}
\label{fig4}
\end{figure}

With the same variational approach, we obtain the Lagrangian
of the system (keeping dynamical variables up to second order)
\begin{eqnarray}
L &=& N\sum\limits_n {\left[ {\frac{{x_n^2}}{2}+\frac{{c_n^2}}{2}
-{x_n}{{\dot c}_n}} \right.}\nonumber \\
&+&\left.
\sum_{|m|=1}^{N_c}{\frac{{N{a_d}{{({x_n} - {x_{n + m}})}^2}}}{4R_x^3 R_y R_z}}
G_{m}\right]
\label{eq:langsecop}
\end{eqnarray}
with $G_{m}$ being just $G$ in Eq.~(\ref{eq:dipoleper})
with $\cos(k_zz_0/R_z)$ term replaced by $\cos(mk_z z_0/R_z)$.
Similar to those in Eq.~(\ref{eq:tfun}),
dynamical variable $x_n=x_n(t)$ corresponds to
the fluctuation of the center of condensate
of layer $n$ in $x$-direction, while
$c_n=c_n(t)$ corresponds to the sloping phases of the condensates of layer $n$.
Assuming that $q_n=q_0\exp(i\omega t-nkz_0)$
($q_n$ represents any one of the dynamical variables in layer $n$ and
$k_z\rightarrow k$ for brevity afterwards),
we obtain the following dispersion relations for the transverse modes:
\begin{eqnarray}
\omega^2(k)=
1 + \frac{Na_d }{R_x^3 R_y R_z} \Lambda(N_c),
\label{eq:discenterop}
\end{eqnarray}
where
\begin{eqnarray}
\Lambda(N_c)\equiv\sum_{|m|=1}^{N_c}\left[1-\cos(mk z_0)\right]G_{m}.
\label{eq:discenterop2}
\end{eqnarray}
The above mode is analogous to an optical phonon mode in crystals to which
$\omega(k=0)$ corresponds to a in-phase mode, while
$\omega(k=\pi/z_0)$ corresponds
to an out-phase mode for neighboring layers.
These two special cases are illustrated
in Fig.~\ref{fig3}. It can be simply checked that $\omega(k=0)=1$. That is,
in-phase mode frequency is just $\omega_x$. Moreover,
if we include only the nearest-neighbors ($N_c=1$)
for $\Lambda$ in Eq.~(\ref{eq:discenterop2}), out-of-phase mode
$\omega(k=\pi/z_0)$ will be exactly the same as $\omega_o$ for the
bilayer system [see Eq.~(\ref{eq:discenter})].
At the long-wavelength limit ($kz_0\ll1$), one obtains
\begin{eqnarray}
\omega^2(k)\simeq 1+ v^2(N_c)k^2,
\label{eq:discenteropsmallk}
\end{eqnarray}
where the square of phonon velocity
\begin{eqnarray}
v^2(N_c)\equiv
\frac{Na_dz_0^2}{R_x^3 R_y R_z}\sum_{|m|=1}^{N_c}{m^2G_{m}}.
\label{eq:discenteropsmallk2}
\end{eqnarray}
Thus $v^2\propto Na_d$ at the low-$k$ limit.
Measurements of dispersion relations of the phonon modes
thus can give direct information on the value of DDI.

In Fig.~\ref{fig4}, we plot the dispersion relations of transverse phonon mode
$\omega$ with $(Na_s,N a_d)=(0,1)$ and $(0.5,1)$ respectively.
The numbers used are based on assuming that
number of $^{52}$Cr atoms in each layer is $N=400$,
magnetic length $\ell$ is about $1\mu$m, and hence $N a_d$ is about $1$.
The curves
presented in Fig.~\ref{fig4} are obtained using a proper
truncation number $N_c$ leading to satisfactory converged results (see later).
Moreover, we assume that $sE_r=300$ and $z_0=0.7 \ell$ and hence $s$ is
about $30$, which is in the deep optical lattice regime.
By minimizing the energy functional (\ref{eq:energyfunopt1}),
we obtain that
$R_z/R\simeq 0.074$ and $z_0/R_z\simeq 7.0$
for the $(Na_s,N a_d)=(0,1)$ case and
$R_z/R\simeq 0.061$ and $z_0/R_z\simeq 7.0$
for the $(Na_s,N a_d)=(0.5,1)$ case.
Inset of Fig.~\ref{fig4} shows the linear dependence of $k^2$ on
$\omega^2$ at the long wavelength limit.
The slope is equal to the square of phonon velocity, $v^2(N_c)$.
When the curve of finite $a_s$ is compared to that of $a_s=0$,
one sees that repulsive
$s$-wave coupling $a_s$ acts to suppress the phonon mode as well as the
phonon velocity. While short-range $a_s$ plays no role between neighboring
layers, its repulsion
actually increases the radii of condensate ($R_x$, $R_y$, $R_z$) in each layer
and consequently
$\Lambda(N_c)/R_x^3 R_y R_z$ becomes smaller.
This indicates that $\omega^2$ as well as the phonon velocity
will be smaller.

\begin{figure}[ptb]
\vspace{-0.0cm}
\includegraphics[width=0.45\textwidth]{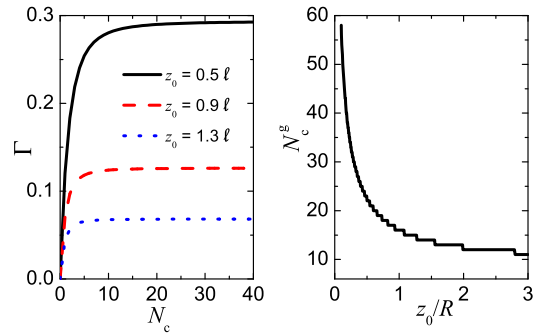}
\vspace{-0.5cm} \caption {(a) $\Gamma(N_c)$ plotted as the function
of $N_c$ for three choices of $z_0=0.5\ell$, $0.9\ell$, and $1.3\ell$ and fixed
$R=1\ell$ and $R_z=0.1\ell$. (b) Critical number $N_c^g$
(see text for definition) plotted as the function of $z_0/R$.}
\label{fig5}
\end{figure}

Finally the behaviors of $\Gamma(N_{c})$, $\Lambda(N_{c})$,
and $v^2(N_{c})$ as the function of $N_c$ are studied.
It is important to first note that $\Gamma(N_{c})$,
$\Lambda(N_{c})$, and $v^2(N_{c})$ all exhibit the same
converging behavior. In Fig.~\ref{fig5}(a),
$\Gamma(N_{c})$ is plotted as the function of $N_{c}$ for
fixed $R_z=0.1\ell$ and $R=1\ell$ and three choices of
$z_0=0.5\ell$, $0.9\ell$, and $1.3\ell$.
It is seen clearly in Fig.~\ref{fig5}(a) that
$\Gamma(N_{c})$ converges at some value of $N_c$
and the larger the $z_0$ is, the smaller the $N_c$ is for the converging result.
To be more explicit, we define a critical value $N_c^g$ of $N_c$ such that
\begin{eqnarray}
{\Gamma(N_c^g)-\Gamma(N_c^g-1)\over\Gamma(N_c^g)}\alt 10^{-3}.
\label{eq:Ncg}
\end{eqnarray}
In fact, $N_{c}^g=19$ and $25$ respectively for the two curves
in Fig.~\ref{fig4}.

Fig.~\ref{fig5}(b) plots $N_c^g$ as a function of $z_0/R$.
It is found that $N_{c}^g$ depends only on the ratio of $z_0/R$
regardless of the value of $R_z$. This occurs because for the
current nonoverlapping multilayer system, $R_z\ll z_0$ and $R_z$
is no longer a well-defined length scale owing to the long-range
character of DDI. As a matter of fact,
$N_c^g$ depends only on the ratio of $z_0/R$.
%This means that for
%$(z_0/R_z, R_z/R)=(5, 1/5)$,
%$(10, 1/10)$, or $(20, 1/20)$ combinations, all of them have
%the same ratio $z_0/R=1$ and will correspond to a same
%critical $N_{c}^g=16$ (see Fig.~\ref{fig5}).
When the ratio $z_0/R$ is larger, the corresponding
$N_c^g$ is smaller and $N_c^g\rightarrow 1$ in the limit of large $z_0/R$.
The results of $N_{c}^g$ indeed can help clarifying whether our results of transverse
phonon modes are trusty or not. How does it work?
Let us consider in a real experiment that
total number of layers is $N_{s}=100$ and the ratio $z_0/R$ is about $1$.
According to the results of Fig.~\ref{fig5}(b), $N_{c}^g=16$ for $z_0/R=1$.
In this case, one does meets the criterion $N_c\ll N_s$ and
our results of transverse
phonon modes are trusty when a real experimental measurement is compared to.

%And we find the convergence speed of these functions is direct proportion
%to $\exp{z_0/R}$.

\section{Conclusions} \label{sec5}
In this paper, analytical solutions of mode frequencies
for the motions of the center of condensates are studied
in bilayer and multilayer (quasi-1D optical lattice) stacks.
In the bilayer, it is
shown that while DDI plays no role
in the in-phase modes of center motions of condensates,
out-of-phase modes ($\omega_o$) depend crucially on the strength of DDI ($a_d$).
More explicitly, $\omega_o$ will depend on
condensate radii of each layer ($R_x,R_y,R_z$), interlayer spacing ($z_0$),
as well as $Na_d$.
Therefore one can actually extract the value of $a_d$
if $\omega_o$, ($R_x,R_y,R_z$), and $z_0$ are measured experimentally.
In the multilayer stack system, transverse (optical) phonon modes
and phonon velocity are derived explicitly. Proper treatment was made
for the boundary effect and it turns out that the truncation number
to which how many neighboring sites should be included for the
long-range DDI is a function of $z_0/R$ ($R$ is the
condensate radius in the transverse direction) only.

\acknowledgements

We are grateful to the support of National Science Council
(Grant No.: NSC 96-2112-M-003-008-MY3) and
National Center for Theoretical Sciences, Taiwan.

%\bibliography{centermotion}
%\bibliographystyle{prsty}

\end{document}